\documentclass[12pt]{article}

\begin{document}

\title{\bf Kinematic Self-Similar Solutions of Locally Rotationally Symmetric Spacetimes}

\author{M. Sharif \thanks{msharif@math.pu.edu.pk} and M. Jamil Amir
\thanks{mjamil.dgk@gmail.com}\\
Department of Mathematics, University of the Punjab,\\
Quaid-e-Azam Campus, Lahore-54590, Pakistan.}

\date{}

\maketitle

\begin{abstract}
This paper contains locally rotationally symmetric kinematic
self-similar perfect fluid and dust solutions. We consider three
families of metrics which admit kinematic self-similar vectors of
the first, second, zeroth and infinite kinds, not only for the
tilted fluid case but also for the parallel and orthogonal cases. It
is found that the orthogonal case gives contradiction both in
perfect fluid and dust cases for all the three metrics while the
tilted case reduces to the parallel case in both perfect fluid and
dust cases for the second metric. The remaining cases give
self-similar solutions of different kinds. We obtain a total of
seventeen independent solutions out of which two are vacuum. The
third metric yields contradiction in all the cases.
\end{abstract}


\section{Introduction}

Due to highly non-linear nature of the Einstein field equations
(EFEs),
\begin{eqnarray}
R_{ab}-\frac{1}{2}g_{ab}R=8\pi G T_{ab},
\end{eqnarray}
the most general analytical solution has not been found till now.
Although, thousands number of solutions are available in the
literature but one has to impose some symmetry restrictions while
solving these equations for any physical system. One of these
symmetry restrictions is called isometry or Killing vector (KV),
which leads to some conservation laws [1]. There has been a recent
literature [2-8, and references therein] which shows a significant
interest in the study of various types of symmetry.

Self-similarity is a powerful tool to simplify the field equations.
In General Relativity (GR), there does not exist any characteristic
scale. A set of field equations remains invariant under a scale
transformation for an appropriate matter field. This means that
there exist scale invariant solutions to the EFEs, known as
self-similar solutions. These solutions often play an important role
in cosmological situations and gravitational collapse. The main
advantage of self-similarity is that it reduces the number of
independent variables by introducing a self-similar variable and
hence reduces the field equations. This variable is a dimensionless
combination of the independent variables, namely the space
coordinates and the time. In other words, self-similarity simply
reduces a system of partial differential equations to an ordinary
differential equations.

In GR, many authors [9] investigated self-similar solutions for
obtaining the realistic solutions of gravitational collapse. There
exist several preferred geometric structures in self-similar models
and one can use different approaches to study these models, such as,
co-moving, homothertic, Schwarzschild approach etc. As pioneers,
Cahill and Taub [10] used the co-moving approach to study
self-similar solutions in the context of cosmology. In this
approach, the coordinates are adopted to the fluid 4-velocity
vector. In GR, self-similarity is defined by the existence of a
homothetic vector (HV) field. Such similarity is called the first
kind (or homothety or continuous self-similarity (CSS)). There
exists a natural generalization of homothety called kinematic
self-similarity defined by the existence of a kinematic self-similar
(KSS) vector field.

Cahill and Taub [10] floated the idea of self-similarity in GR. They
showed that it corresponds to self-similarity of the homothetic
class in the context of Newtonian theory and is known as KSS of the
first kind. Later, self-similarity of the second, zeroth and
infinite kinds were introduced by Carter and Henriksen [11,12].
There is a great literature available [2,3,13-18] which contain
several KSS perfect fluid solutions of the EFEs. It has been shown
that $p=k\rho$ is the only barotropic equation of state which is
compatible with self-similarity of the first kind.  Carr [2]
classified the most general spherically symmetric dust solutions of
the EFEs admitting the self-similarity of the first kind. This work
was extended by Carr and Coley [3] for all spherically symmetric
perfect fluid solutions. Coley [13] concluded that the solutions in
which the KSS vector is parallel to the fluid flow are necessarily
Friedmann-Robertson-Walker models. McIntosh [14] proved that a
vacuum spacetime admits only a non-trivial homothetic motion if the
homothetic vector field is non-null and is not hypersurface
orthogonal. Benoit and Coley [15] studied spherically symmetric
perfect fluid solutions of the EFEs admitting KSS vector of the
first, second and zeroth kinds by using analytic approach.

Sintes et al. [16] investigated the perfect fluid solutions in the
case of self-similarity of the infinite kind. Carr et al. [17]
considered the KSS solutions associated with the critical behavior
observed in the gravitational collapse of spherically symmetric
perfect fluid by using the equation of state $p=k\rho$. They also
investigated solution space of self-similar spherically symmetric
perfect fluid models [18] and discussed the physical aspects of
these solutions. Coley and Goliath [19] studied the self-similar
spherically symmetric cosmological models with a perfect fluid and a
scalar field by using an exponential potential.

Maeda et al. [4,5] analyzed the KSS solution of the second kind for
the titled perfect fluid case by using a relativistic polytropic
equation of state. They classified the spherically symmetric perfect
fluid and dust solutions admitting the KSS vector of different kinds
[6] and found some interesting solutions. Sharif and Sehar [20]
extended this work for cylindrically symmetric spacetimes for both
perfect fluid and dust cases with tilted, parallel and orthogonal
vector fields by using different equations of state. They also
studied the physical properties of spherically [21], cylindrically
[22] and plane [23] symmetric spacetimes.

Recently, Sharif and Sehar [24,25] have explored the KSS solutions
of the most general plane symmetric spacetimes. Sintes [26] explored
some KSS solutions of locally rotationally symmetric (LRS)
spacetimes. This paper is devoted to complete the study of the KSS
solutions of LRS models both for the perfect fluid and dust cases.
The kinematic self-similar vectors of the first, second, zeroth and
infinite kinds for the tilted fluid as well as parallel and
orthogonal cases would be investigated.

The paper is organized as follows. In section 2, we shall give three
metrics representing  the non-static LRS spacetimes. Section 3 is
devoted to find the KSS perfect fluid and dust solutions of
different kinds for the first LRS metrics. In section 4, we shall
discuss all possible KSS solutions of the second LRS metrics. The
last section contains a summary and discussion of the results
obtained.

\section{Locally Rotationally Symmetric Models and Kinematic Self-Similarity}

Many authors [27-30] studied extensively the LRS spacetimes which
contain well-known exact solutions of the EFEs. They admit a group
of motions $G_4$ acting multiply transitively on three dimensional
non-null orbits spacelike ($S_3$) or timelike ($T_3$) and the
isotropy group is a spatial rotation. These spacetimes are
represented by three families of metrics given as [26,27]
\begin{eqnarray}
ds^2&=&\epsilon[-dt^2+ A^2(t)dx^2]-B^2(t)dy^2-B^2(t)\Sigma^2(y,k)dz^2,\\
ds^2&=&\epsilon[-dt^2+ A^2(t)\{dx-\Lambda(y,k)dz\}^2]-B^2(t)dy^2\nonumber \\
&-&B^2(t)\Sigma^2(y,k)dz^2,\\
ds^2&=&\epsilon[-dt^2+ A^2(t)dx^2]-B^2(t)dy^2-B^2(t)\Sigma^2(y,k)dz^2,
\end{eqnarray}
where $k=-1,0,1,~\epsilon=\pm1$,
\begin{eqnarray}
\Sigma=\left\{\begin{array}{l}
\sin y,  \quad~~ k=1,\\
y, \quad\quad\quad k=0,\\
\sinh y, \quad k=-1,\\
\end{array}\right.
\end{eqnarray}
and
\begin{eqnarray}
\Lambda=\left\{\begin{array}{l}
\cos y,  \quad~~ k=1,\\
\frac{y^2}{2}, \quad\quad~~ k=0,\\
\cosh y, \quad k=-1.\\
\end{array}\right.
\end{eqnarray}
The static and non-static solutions correspond to $\epsilon=1$ and
$\epsilon=-1$ respectively. We restrict our attention towards the
non-static case as the results for the static case can be obtained
consequently.  For $\epsilon=-1$, the above equations take the form
\begin{eqnarray}
ds^2&=&dt^2- A^2(t)dx^2-B^2(t)dy^2-B^2(t)\Sigma^2(y,k)dz^2,\\
ds^2&=&dt^2- A^2(t)dx^2-B^2(t)e^{2x}dy^2-B^2(t)e^{2x}dz^2,\\
ds^2&=&dt^2-A^2(t)dx^2-B^2(t)dy^2-\{A^2(t)\Lambda^2(y,k)\nonumber\\
&+&B^2(t)\Sigma^2(y,k)\}dz^2+2A^2(t)\Lambda(y,k)dxdz.
\end{eqnarray}
The metrics (7) become Bianchi types $I(BI)$ or $VII_0$ $(BVII_0)$
for $k=0$, $III$ $(BIII)$ for $k=-1$ and Kantowski-Sachs (KS) for
$k=+1$. The metrics (8) represent Bianchi type $V (BV)$ or $VII_h$
$(BVII_h)$ metric. The metrics (9) turn out to be Bianchi types $II
(BII)$ for $k=0$, $VIII (BVIII)$ or $III (BIII)$ for $k=-1$ and $IX
(BIX)$ for $k=+1$. The energy-momentum tensor for a perfect fluid is
given as
\begin{equation}
T_{ab}=[\rho(t,y)+p(t,y)]u_{a}u_{b}+ p(t,y )g_{ab},\quad
(a,b=0,1,2,3),
\end{equation}
where $\rho$ is the density, $p$ is the pressure and $u_{a}$ is the
four-velocity of the fluid element in the co-moving coordinate
system given as $u_{a}=(1,0,0,0)$.

A kinematic self-similar vector $\xi$ satisfies the following two
conditions
\begin{eqnarray}
\pounds_{\xi}h_{ab}&=& 2\delta h_{ab},\\
\pounds_{\xi}u_{a}&=& \alpha u_{a},
\end{eqnarray}
where $h_{ab}=g_{ab}-u_au_b$ is the projection tensor, $\alpha$ and
$\delta$ are constants. The similarity transformation is
characterized by the scale independent ratio, $\alpha/\delta$, known
as the similarity index which yields the following two cases:
\begin{eqnarray*}
1.\quad\delta\neq0;\quad 2.\quad \delta=0.
\end{eqnarray*}
The case 1 yields self-similarity of the first, zeroth, second kinds
and the case 2 corresponds to self-similarity of the infinite kind
respectively. These are discussed below in detail.

\section{Kinematic Self-similar Solutions of the First Metric}

For this metric, the EFEs reduce to the following form
\begin{eqnarray}
\frac{2\dot{A}\dot{B}}{AB}+\frac{\dot{B}^2}{B^2}+\frac{K}{B^2}&=&\kappa \rho,\\
\frac{\ddot{A}}{A}+\frac{3\ddot{B}}{B}+\frac{\dot{B}^2}{B^2}+\frac{\dot{A}\dot{B}}
{AB}+\frac{K}{B^2}&=&-2\kappa
p,
\end{eqnarray}
where dot represents the differentiation w.r.t. time coordinate. The
conservation of energy-momentum tensor ${T^{ab}}_{;b}=0$ yields the
following equation
\begin{eqnarray}
\dot{\rho}+(\frac{\dot{A}}{A}+\frac{2\dot{B}}{B})(p+\rho)=0, \quad
p'=0,
\end{eqnarray}
where prime means differentiation w.r.t. $y$. In this case, the
general form of a vector field $\xi$ may be given as
\begin{equation}
\xi^{a}\frac{\partial}{\partial
x^a}=h_{0}(t,y)\frac{\partial}{\partial
t}+h_{2}(t,y)\frac{\partial}{\partial y},
\end{equation}
where $h_{0}$ and $h_{2}$ are arbitrary functions. When $\xi$ is
parallel to the fluid flow, $h_{2}=0$ while $h_{0}=0$ indicates
that $\xi$ is orthogonal to the fluid flow. When both $h_{0}$ and
$h_{2}$ are non-vanishing, $\xi$ will be tilted to the fluid flow.

\subsection{Perfect Fluid Case}

Here we discuss all the above kinds when the vector field is tilted,
parallel and orthogonal to the fluid flow for the perfect fluid.

\subsubsection{Tilted Vector Field}

\textbf{Case 1:} For the sake of simplicity we choose $\delta$ as
unity, then the KSS equations will become
\begin{eqnarray}
{\xi^0}_{,0}=\alpha,\\
{\xi^0}_{,2}-B^2{\xi^2}_{,0}=0,\\
\frac{\dot{A}}{A}\xi^0=1,\\
\frac{\dot{B}}{B}\xi^0+{\xi^2}_{,2}=1,\\
\frac{\dot{B}}{B}\xi^0+\frac{\Sigma'}{\Sigma}{\xi^2}_{,2}=1.
\end{eqnarray}
From Eqs.(17) to (21), we obtain the following form of $\xi^0$ and
$\xi^2$
\begin{equation}
\xi^{a}\frac{\partial}{\partial x^a}=(\alpha
t+\beta)\frac{\partial}{\partial t}+
c\Sigma\frac{\partial}{\partial y}.
\end{equation}
For the tilted case, the similarity index, $\alpha/\delta$, yields
the following three different possibilities
\begin{eqnarray*}
&(i)&\quad\delta\neq0,\quad\alpha=1\quad
(\beta~\rm{can~ be~ taken~ to~ be~ zero}),\\
&(ii)&\quad\delta\neq0,\quad\alpha=0\quad
(\beta~\rm{can~ be~ taken~ to~ be~ unity}),\\
&(iii)&\quad\delta\neq0,\quad\alpha\neq0,1\quad (\beta~\rm{can~ be
taken~ to~ be~ zero}).
\end{eqnarray*}
These cases correspond to the self-similarity of {\it first}, {\it
zeroth} and {\it second kind} respectively. We shall discuss these
kinds separately.

For self-similarity of the first kind, we have
$\xi^0=t,~\xi^2=c\Sigma$. The corresponding solution for $k=0$ takes
the form
\begin{eqnarray}
ds^2&=&dt^2-a^2t^2dx^2-b^2t^{2(1-c)}(dy^2+y^2dz^2),\nonumber\\
\rho&=&\frac{9}{16\kappa t^2}=-3p,
\end{eqnarray}
where $a$ and $b$ are arbitrary constants but $c=\frac{1}{2},1,2$.
There exists no solution for $k=\pm 1$.

Self-similarity of the zeroth kind yields $\xi^0=1,~\xi^2=c\Sigma$
and the solution for $k=0$ turns out to be
\begin{eqnarray}
ds^2&=&dt^2-a^2e^{2t}dx^2-b^2e^{2(1-c)t}(dy^2+y^2dz^2),\nonumber\\
\rho&=&\frac{1}{\kappa}(c-1)(c-3), \quad
p=\frac{-1}{2\kappa}(4c^2-9c+6).
\end{eqnarray}
Here $c=0,\frac{1}{2}, \frac{3}{2}$ and no solution exists for
$k=\pm 1$.

In the second kind, $\xi^0=\alpha t,~\xi^2=c\Sigma$ and the
corresponding solution for $k=0$ becomes
\begin{eqnarray}
ds^2&=&dt^2-a^2t^{2/\alpha}dx^2-b^2t^{2(1-c)/\alpha}(dy^2+y^2dz^2),\nonumber\\
\rho&=&\frac{1}{\kappa \alpha^2 t^2}(c-1)(c-3),\nonumber\\
p&=&\frac{-1}{2\kappa \alpha^2 t^2}(4c^2-9c+6+3\alpha c-4\alpha),
\end{eqnarray}
where $c$ is either zero or $\frac{1}{2},\frac{3-\alpha}{2}$. There
does not exist any solution for $k=\pm 1$.\\
\par\noindent
\textbf{Case 2:} For $\delta=0$ and $\alpha\neq0$ ($\alpha$ can be
unity and $\beta$ can be re-scaled to zero), the self-similarity is
known as {\it infinite kind}. In this case, the KSS equations take
the following form
\begin{eqnarray}
{\xi^0}_{,0}=1,\\
{\xi^0}_{,2}-B^2{\xi^2}_{,0}=0,\\
\frac{\dot{A}}{A}\xi^0=0,\\
\frac{\dot{B}}{B}\xi^0+{\xi^2}_{,2}=0,\\
\frac{\dot{B}}{B}\xi^0+\frac{\Sigma'}{\Sigma}{\xi^2}_{,2}=0.
\end{eqnarray}
Here $\xi^0=t+c_0,~\xi^2=c\Sigma$ and the corresponding solution for
$k=0$ is
\begin{eqnarray}
ds^2&=&dt^2-a^2dx^2-b^2(t+c_0)^{-2c}(dy^2+y^2dz^2),\nonumber\\
\rho&=&\frac{c^2}{\kappa(t+c_0)^2},\quad p=\frac{-c(4c+3)}{2\kappa
(t+c_0)^2},
\end{eqnarray}
where $c=0,-\frac{1}{2}$. It is mentioned here that the solutions
become vacuum and stiff fluid for $c=0$ and $c=-\frac{1}{2}$
respectively. The solution for $k=+1$ is given by
\begin{eqnarray}
ds^2&=&dt^2-a^2dx^2-b^2(dy^2+\sin^2ydz^2),\nonumber\\
\rho&=&-p/2.
\end{eqnarray}
The solution for $k=-1$ is
\begin{eqnarray}
ds^2&=&dt^2-a^2dx^2-b^2(dy^2+\sinh^2ydz^2),\nonumber\\
\rho&=&-p/2.
\end{eqnarray}
It is noted that $\xi^2$ vanishes for $k=\pm 1$, i.e., these
solutions fall in the parallel case.

\subsubsection{Parallel Vector Field }

\textbf{Case 1:} In this case $\xi^1,\xi^2,\xi^3=0$ and the KSS
equations for $\delta=1$ take the form
\begin{eqnarray}
{\xi^0}_{,0}=\alpha,\\
\frac{\dot{A}}{A}\xi^0=1,\\
\frac{\dot{B}}{B}\xi^0=1.
\end{eqnarray}
Integrating Eq.(34), we obtain
\begin{equation}
\xi^0=(\alpha t+\beta).
\end{equation}

For self-similarity of the first kind, $\xi^0=t$ and the
corresponding solution for $k=0$ becomes
\begin{eqnarray}
ds^2&=&dt^2-a^2t^2dx^2-b^2t^2(dy^2+y^2dz^2),\nonumber\\
\rho&=&\frac{3}{\kappa t^2}=-3p.
\end{eqnarray}

In the zeroth kind, $\xi^0=1$ and the corresponding solution for
$k=0$ is
\begin{eqnarray}
ds^2&=&dt^2-a^2e^{2t}dx^2-b^2e^{2t}(dy^2+y^2dz^2),\nonumber\\
\rho&=&\frac{3}{\kappa}=-p.
\end{eqnarray}

Self-similarity of the second kind, $\xi^0=\alpha t$, yields the
following solution for $k=0$
\begin{eqnarray}
ds^2&=&dt^2-a^2t^{2/\alpha}dx^2-b^2t^{2/\alpha}(dy^2+y^2dz^2),\nonumber\\
\rho&=&\frac{3}{\kappa \alpha^2 t^2},\quad p=-\frac{1}{\kappa
\alpha^2 t^2}(3-2\alpha).
\end{eqnarray}
There exists no solution for $k=\pm 1$ in the above kinds.\\
\par\noindent
\textbf{Case 2:} The infinite kind, $\delta=0,~\alpha\neq0$ yields
the same solution as given in Eq.(31) for $c=0$. The solution for
$k=\pm 1$ are the same as given by Eqs.(32) and (33).

\subsubsection{Orthogonal Vector Field}

There does not exist any solution for the perfect fluid case when
the vector field is orthogonal to the fluid flow.

\subsection{Dust Case}

For the dust case, we take $p=0$  in Eqs.(14) and (15) so that
\begin{eqnarray}
\frac{\ddot{A}}{A}+\frac{3\ddot{B}}{B}+\frac{\dot{B}^2}{B^2}+
\frac{\dot{A}\dot{B}}{AB}+\frac{K}{B^2}&=&0,\\
\dot{\rho}+(\frac{\dot{A}}{A}+\frac{2\dot{B}}{B})\rho=0,
\end{eqnarray}
The KSS solutions for the tilted case are given in table 1

\vspace{0.5cm}

{\bf {\small Table 1.}} {\small KSS solutions for the tilted dust
case when $k=0$}

\vspace{0.5cm}

\begin{center}
\begin{tabular}{|l|l|l|l|l|l|}
\hline {\bf Kinds}& {\bf A(t)}& {\bf B(t)}& {\bf Density}& {\bf KSS Vectors}\\
\hline First Kind& $ at$&$ b,~bt^{1/2}$&$ 0$&$ \xi^0=t, \quad \xi^2=y$ \\
\hline Zeroth Kind & $-$&$-$&$-$&$-$\\
\hline Second Kind & $ at^{-\frac{1}{3}}$&$ bt^{\frac{2}{3}}$&$ 0$&$
\xi^0=-3t, \quad \xi^2=3y$ \\
\hline Infinite kind & $ a$&$ b$&$ 0$&$ \xi^0=t+c_0, \quad \xi^2=0$ \\
\hline
\end{tabular}
\end{center}
Here $a$, $b$ and $c_0$ are arbitrary constants. The solution for
the infinite kind will fall in the parallel dust case as $\xi^2$
vanishes for this case. The KSS solutions for the parallel dust
case, when $\xi^2=0$, are given in table 2

\vspace{0.5cm}

{\bf {\small Table 2.}} {\small KSS solutions for the parallel dust
case when $k=0$}

\vspace{0.5cm}

\begin{center}
\begin{tabular}{|l|l|l|l|l|l|}
\hline {\bf Kinds}& {\bf A(t)}& {\bf B(t)}& {\bf Density}& {\bf KSS Vectors}\\
\hline First Kind& $ -$&$ -$&$ -$&$ -$\\
\hline Zeroth Kind& $-$&$ -$&$ -$&$ -$\\
\hline Second Kind & $ at^{\frac{2}{3}}$&$ bt^{\frac{2}{3}}$&$ \frac{4}
{3\kappa t^2}$&$ \xi^0=\frac{3}{2}t$ \\
\hline Infinite kind & $ a$&$ b$&$ 0$&$ \xi^0=t+c_0$ \\
\hline
\end{tabular}
\end{center}
There exists no solution in all the cases mentioned in the tables 1
and 2 for $k=\pm 1$. Further, there exists no solution for the dust
case when the vector field is orthogonal to the fluid flow.

\section{Kinematic Self-similar Solutions of the 2nd Metric}

For this metric (8), the EFEs reduce to the following form
\begin{eqnarray}
\frac{2\dot{A}\dot{B}}{AB}+\frac{\dot{B}^2}{B^2}-\frac{3}{A^2}&=&\kappa \rho,\\
\frac{\ddot{A}}{A}+\frac{3\ddot{B}}{B}+\frac{\dot{B}^2}{B^2}+\frac{\dot{A}
\dot{B}}{AB}-\frac{2}{A^2}&=&-2\kappa
p,\\
\frac{\dot{A}}{A}-\frac{\dot{B}}{B}=0.
\end{eqnarray}
The conservation of energy-momentum tensor ${T^{ab}}_{;b}=0$ yields
the same equation as given by Eq.(16). Here prime means
differentiation w.r.t. $x$. For these spacetimes, the general form
of a vector field $\xi$ may be given as
\begin{equation}
\xi^{a}\frac{\partial}{\partial
x^a}=h_{0}(t,x)\frac{\partial}{\partial
t}+h_{1}(t,x)\frac{\partial}{\partial x},
\end{equation}
where $h_{0}$ and $h_{1}$ are arbitrary functions.

\subsection{Perfect Fluid Case}

Now we shall discuss the cases (1) and (2) when the vector field is
tilted, parallel and orthogonal to the fluid flow.

\subsubsection{Tilted Vector Field}

This case reduces to the parallel perfect fluid case.

\subsubsection{Parallel Vector Field}

\textbf{Case 1:} For the parallel vector field, we take $\xi^1=0$
and $\delta=1$ and the KSS equations imply that $\xi^0=\alpha
t+\beta$. For the self-similarity of the first kind, $\xi^0=t$ and
the corresponding solution take the form
\begin{eqnarray}
ds^2&=&dt^2-a^2t^2dx^2-b^2t^2e^{2x}(dy^2+dz^2),\nonumber\\
\rho&=&-\frac{3}{\kappa t^2}(\frac{1}{a^2}-1)=-3p.
\end{eqnarray}

In self-similarity of the zeroth kind, $\xi^0=1$ and the
corresponding solution is
\begin{eqnarray}
ds^2&=&dt^2-a^2e^{2t}dx^2-b^2e^{2(t+x)}(dy^2+dz^2),\nonumber\\
\rho&=&\frac{3}{\kappa}(1-\frac{1}{a^2}~e^{-2t})=-p.
\end{eqnarray}

For the second kind, $\xi^0=\alpha t$, the solution becomes
\begin{eqnarray}
ds^2&=&dt^2-a^2t^{2/\alpha}dx^2-b^2t^{2/\alpha}e^{2x}(dy^2+dz^2),\nonumber\\
\rho&=&\frac{3}{\kappa}(\frac{1}{a^2t^2}-\frac{1}{a^2t^{\frac{2}{\alpha}}}),\quad
p=-\rho+\frac{2}{\kappa\alpha t^2}.
\end{eqnarray}
\textbf{Case 2:} The infinite kind, $\delta=0,~\alpha\neq0$, leads
to $\xi^0=t+c_0,~A=a,~ B=b$ and the corresponding solution turns out
to be
\begin{eqnarray}
ds^2&=&dt^2-a^2dx^2-b^2e^{2x}(dy^2+dz^2),\nonumber\\
\rho&=&-\frac{3}{\kappa a^2}=-3p.
\end{eqnarray}

\subsubsection{Orthogonal Vector Field }

Here all the possibilities lead to contradiction.

\subsection{Dust Case}

For the dust case, we take $p=0$ in Eqs.(14) and (15) so that
\begin{eqnarray}
\frac{\ddot{A}}{A}+\frac{3\ddot{B}}{B}+\frac{\dot{B}^2}{B^2}+
\frac{\dot{A}\dot{B}}{AB}+\frac{K}{B^2}&=&0,\\
\dot{\rho}+(\frac{\dot{A}}{A}+\frac{2\dot{B}}{B})\rho=0,
\end{eqnarray}
The tilted dust case reduces to the parallel dust case. The KSS
solutions for the parallel case, when $\xi^1=0$, are given in table
3.

\vspace{0.5cm}

{\bf {\small Table 3.}} {\small KSS solutions for the parallel dust
case}

\vspace{0.5cm}

\begin{center}
\begin{tabular}{|l|l|l|l|l|l|}
\hline {\bf Kinds}& {\bf A(t)}& {\bf B(t)}& {\bf Density}& {\bf KSS Vectors}\\
\hline First Kind& $ \pm 1$&$ b$&$ 0$&$ \xi^0=t$\\
\hline Zeroth Kind & $ -$&$ -$&$-$&$ -$\\
\hline Second Kind & $ -$&$ -$&$-$&$ -$ \\
\hline Infinite kind & $ -$&$ -$&$-$&$ -$\\
\hline
\end{tabular}
\end{center}
The dust orthogonal case yields contradiction for all the
possibilities. Also, we obtain contradiction in all the cases of the
third metric (9).

\section{Conclusion}

We have investigated the LRS spacetimes which admit self-similarity
of the first, zeroth, second and infinite kinds for both perfect
fluid and dust cases. We have explored the possibilities when KSS
vector is tilted, parallel or orthogonal to the fluid flow. We find
a total of seventeen independent KSS solutions out of which two are
vacuum.

For the metric (7), there arise three KSS solutions in the tilted
perfect fluid case and coincide with the results given by Sintes
[25] for $n=0$ and $m=c$ in the first, zeroth and second kinds. For
the infinite kind, we find three solutions which do not agree with
the solutions given in [25]. It is mentioned here that the KSS
solutions of the first kind turns out to be the radiation case and
the tilted perfect fluid case of the infinite kind reduces to the
parallel perfect fluid case for $k=\pm 1$. The parallel perfect
fluid case gives three independent KSS solutions in the first,
zeroth and second kinds. These solutions also coincide with those
given in [25]. The infinite kind yields the same solutions as for
the tilted perfect fluid case of the infinite kind when $c=0$. For
the tilted dust case, we have two KSS solutions of the first kind
and one of the second kind which are vacuum. The infinite kind gives
the same solution as given in Eq.(31) for $c=0$ while the zeroth
kind gives no solution. In the parallel dust case, the first and
zeroth kinds give contradiction while the infinite kind gives the
same solution as given in Eq.(31) for $c=0$. We obtain one
independent solution in the second kind. The orthogonal case yields
contradiction both in prefect fluid and dust cases.

For the metric (8), the tilted perfect fluid case reduces to the
parallel case and this yields four KSS solutions of the first,
zeroth, second and infinite kinds which coincide with [25] when
$n=0$ and $m=c$. In the dust case, there exists only one KSS
solution of the first kind which coincides with Eq.(50) for $a=\pm1$
while the second, zeroth and infinite kinds yield contradiction. It
is mentioned here that the orthogonal case always gives
contradiction.

For the metric (9), we have contradiction in all the cases. The
summary of the results is given in the following tables.

\vspace{0.2cm}

{\bf {\small Table 4.}} {\small Perfect fluid kinematic self-similar
solutions of the metric (7)}

\vspace{0.1cm}

\begin{center}
\begin{tabular}{|l|l|}
\hline {\bf Self-similarity} & {\bf Solution}
\\ \hline First kind (tilted) & Solution given by
Eq.(23)
\\ \hline First kind (parallel) & Solution given by
Eq.(38)
\\ \hline First kind (orthogonal) & None
\\ \hline Zeroth kind (titled) & Solution given by
Eq.(24)
\\ \hline Zeroth kind (parallel) &Solution given by
Eq.(39)
\\ \hline Zeroth kind (orthogonal) & None
\\ \hline Second kind (tilted) & Solution given by
Eq.(25)
\\ \hline Second kind (parallel) & Solution given by
Eq.(40)
\\ \hline Second kind (orthogonal) & None
\\ \hline Infinite kind (tilted) &  Solution given by
Eq.(31)
\\ \hline Infinite kind (parallel) & Solution given by
Eqs.(33),(31) and \\&(32) for $k=-1,0,1$ respectively
\\ \hline Infinite kind (orthogonal) & None
\\ \hline
\end{tabular}
\end{center}
\newpage

{\bf {\small Table 5.}} {\small Dust kinematic self-similar
solutions of the metric (7)}

\vspace{0.1cm}

\begin{center}
\begin{tabular}{|l|l|}
\hline {\bf Self-similarity} & {\bf Solution}
\\ \hline First kind (tilted) & Two vacuum solutions given in table 1
\\ \hline First kind (parallel) & None
\\ \hline First kind (orthogonal) & None
\\ \hline Zeroth kind (titled) & None
\\ \hline Zeroth kind (parallel) &None
\\ \hline Zeroth kind (orthogonal) & None
\\ \hline Second kind (tilted) & Vacuum solution given in table 1
\\ \hline Second kind (parallel) & Solution given in table 2
\\ \hline Second kind (orthogonal) & None
\\ \hline Infinite kind (tilted) & Vacuum solution given by
Eq.(31)\\& for $c=0$
\\ \hline Infinite kind (parallel) &Vacuum solution given by Eq.(31)
\\& for $c=0$
\\ \hline Infinite kind (orthogonal) & None
\\ \hline
\end{tabular}
\end{center}

\vspace{0.2cm}

{\bf {\small Table 6.}} {\small Perfect fluid kinematic self-similar
solutions of the metric (8)}

\vspace{0.1cm}

\begin{center}
\begin{tabular}{|l|l|}
\hline {\bf Self-similarity} & {\bf Solution}
\\ \hline First kind (tilted) & None
\\ \hline First kind (parallel) & Solution given by
Eq.(47)
\\ \hline First kind (orthogonal) & None
\\ \hline Zeroth kind (titled) & None
\\ \hline Zeroth kind (parallel) &Solution given by
Eq.(48)
\\ \hline Zeroth kind (orthogonal) & None
\\ \hline Second kind (tilted) & None
\\ \hline Second kind (parallel) & Solution given by
Eq.(49)
\\ \hline Second kind (orthogonal) & None
\\ \hline Infinite kind (tilted) & None
\\ \hline Infinite kind (parallel) & Solution given by
Eq.(50)
\\ \hline
\end{tabular}
\end{center}
We would like to mention here that Eqs.(38), (39) and (40) are the
special cases of Eqs.(23), (24) and (25) respectively for $c=0$.
Further, Eqs.(23) and (24) represent orthogonal spatially
homogeneous perfect fluid Bianchi $I$ models with homothetic vector
field and the equation of state with $\gamma=\frac{2}{3}$ and
$\gamma=0$ respectively. These can correspond to FRW models under
particular coordinate transformations. Similarly, Eqs.(47) and (48)
represent the orthogonal spatially homogeneous perfect fluid Bianchi
$V$ models with homothetic vector field and the equation of state
with $\gamma=\frac{2}{3}$ and $\gamma=0$ respectively. Also, Eq.(31)
represents Minkowski space for $c=0$. The dust parallel case of the
first kind yields vacuum solution as a special case given by
Eq.(50).

\newpage
\begin{description}
\item  {\bf Acknowledgment}
\end{description}

We appreciate and acknowledge the Higher Education Commission
Islamabad, Pakistan for its financial support through the {\it
Indigenous PhD 5000 Fellowship Program Batch-I}.
\vspace{1cm}

{\bf \large References}

\begin{description}

\item{[1]} Noether, E. Nachr, Akad. Wiss. Gottingen, II, Math.
           Phys. {\bf K12}(1918)235;\\
           Davis, W.R. and Katzin, G.H. Am. J. Phys. {\bf 30}(1962)750;\\
           Petrov, A.Z.: \textit{Einstein Spaces}
           (Pergamon, Oxford University Press, 1969);\\
           Hojman, L.
           Nunez, l. Patino, A. and Rago, H.: J. Math. Phys. {\bf
           27}(1986)281.

\item{[2]} Carr, B.J.: Phys. Rev. {\bf D62}(2000)044022.

\item{[3]} Carr, B. J. and Coley, A. A.: Phys. Rev. {\bf D62}(2000)044023.

\item{[4]} Maeda, H., Harada, T., Iguchi, H. and Okuyama, N.: Phys. Rev.
           {\bf D66}(2002)027501.

\item{[5]} Maeda, H., Harada, T., Iguchi, H. and Okuyama, N.: Prog. Theor. Phys.
           {\bf108}(2002)819.

\item{[6]} Maeda, H., Harada, T., Iguchi, H. and Okuyama, N.: Prog. Theor. Phys.
           {\bf110}(2003)25.

\item{[7]} Sharif, M.: J. Math. Phys. {\bf 44}(2003)5141; ibid {\bf
           45}(2004)1518; ibid {\bf 45}(2004)1532.

\item{[8]} Sharif, M. and Aziz, S.: Int. J. Mod. Phys. {\bf
D14}(2005)1527.

\item{[9]} Penston, M.V.: Mon. Not. R. Astr. Soc. {\bf144}(1969)425;\\
           Larson, R.B.: Mon. Not. R. Astr. Soc. {\bf145}(1969)271;\\
           Shu, F.H.: Astrophys. J. {\bf214}(1977)488;\\
           Hunter, C.: Astrophys. J. {\bf218}(1977)834.

\item{[10]} Cahill, M. E. and Taub, A. H.:   Commun. Math. Phys. {\bf21}(1971)1.

\item{[11]} Carter, B. and Henriksen, R. N.:  Annales De Physique {\bf14}(1989)47.

\item{[12]} Carter, B. and Henriksen, R. N.:  J. Math. Phys. {\bf32}(1991)2580.

\item{[13]} Coley, A. A.: Class. Quantum Grav. {\bf14}(1997)87.

\item{[14]} McIntosh, C.B.G.: Gen. Rel. Gravit. {\bf7}(1975)199.

\item{[15]} Benoit, P. M. and Coley, A. A.: Class. Quantum Grav. {\bf15}(1998)2397.

\item{[16]} Sintes, M.A., Benoit, P.M. and Coley, A. A.: Gen. Relativ. Gravit. {\bf33}(2001)1863.

\item{[17]} Carr, B. J., Coley, A. A., Golaith, M., Nilsson, U. S. and Uggla, C.:
            Class. Quantum Grav. {\bf18}(2001)303.

\item{[18]} Carr, B. J., Coley, A. A., Golaith, M., Nilsson, U. S. and Uggla, C.:
            Phys. Rev. {\bf D61}(2000)081502.

\item{[19]} Coley, A. A. and Golaith, M.: Class. Quantum Grav. {\bf17}(2000)2557.

\item{[20]} Sharif, M. and Aziz, S.: Int. J. Mod. Phys. {\bf D14}
            (2005)1527.

\item{[21]} Sharif, M. and Aziz, S.: Int. J. Mod. Phys. {\bf D14}
            (2005)73.
\item{[22]} Sharif, M. and Aziz, S.: Int. J. Mod. Phys. {\bf A20}(2005)7579.

\item{[23]} Sharif, M. and Aziz, S.: J. Korean Phys. Soc. {\bf 47}(2005)757.

\item{[24]} Sharif, M. and Aziz, S.: J. Korean Phys. Soc. {\bf 49}(2006)21.

\item{[25]} Sharif, M. and Aziz, S.: Class. Quantum Grav. {\bf 24}(2007)605.

\item{[26]} Sintes, M.A.: Class. Quantum Grav. {\bf 15}(1998)3689.

\item{[27]} Ellis, G.F.R.: J. Math. Phys. {\bf 8}(1967)1171.

\item{[28]} Steward, J.M. and Ellis, G.F.R.: J. Math. Phys. {\bf 9}(1968)1072.

\item{[29]} Ellis, G.F.R. and MacCallum, M.A.H.: Commun. Math. Phys. {\bf 12}(1969)108.

\item{[30]} Tsamparlis, M. and Apostolopoulos, P.S.: Gen. Relativ. Gravit. {\bf36}(2004)47.

\end{description}
\end{document}